# Development of an Ontology Based Forensic Search Mechanism: Proof of Concept

**Jill Slay and Fiona Schulz**
Enterprise Security Management Lab
Advanced Computing Research Centre
School of Computer and Information Science
University of South Australia
Mawson Lakes SA5095 Australia

## ABSTRACT

This paper examines the problems faced by Law Enforcement in searching large quantities of electronic evidence. It examines the use of ontologies as the basis for new forensic software filters and provides a proof of concept tool based on an ontological design. It demonstrates that efficient searching is produced through the use of such a design and points to further work that might be carried out to extend this concept.

## 1. INTRODUCTION

The need to extract evidentiary data from computers has given rise to the discipline of computer forensics. Computer forensics is the name given to the "process of identifying, preserving, analysing and presenting digital evidence in a manner that is legally acceptable" (McKemmish, 1999). Forensic Computing is an area rapidly expanding because of necessity since many crimes are now committed entirely electronically. Instances of computer related crime are on the rise around the world. The advent of broadband Internet has made many such crimes easier, or at the very least faster, to commit.

Law enforcement agencies are faced with an exponential growth in the amount of electronic evidence which needs to be collected from digital devices when found at traditional and computer based crime scenes as well as the more common need to collect evidence of electronic crime, such as that involving fraud or pornography.

Examples (Beckett, 2005) of the growth in electronic evidence which needs investigation are:

**1995 New South Wales Police Evidence**

- 30 Cases per year (Hard Disk Drive Size 10-20MB);
- 40 Computers examined;

- approximately 1.2GB of Data for the year;
- 5 Phone examinations for the year.

**2000 - New South Wales Police Evidence**

- 280 cases per year (Hard Disk Drive Size 1-10GB);
- 200 Computers examined;
- 2TB of data for the year;
- 60 Mobile phone examinations.

**2005 - New South Wales Police Evidence**

- over 500 accepted cases per year after with job acceptance policy (Hard Disk Drive size 40-120GB);
- more than 400TB to date processed;
- more than 1000 computers examined;
- more than 300 Phones examined.

72 different types of device are listed as being currently available (Mohay, 2005), with ever-growing storage capacity, and each of these is able to provide electronic evidence for law enforcement and intelligence purposes.

The use of computers in crime is indisputable, but locating the data that actually relates to a crime is a challenge for investigators. Data may be hard to locate or even have been deliberately hidden or encrypted. One of the primary problems for investigators is thoroughness versus time. An investigation must be completed within a "reasonable period of time" (Anderson, 2001) but sifting through hundreds of gigabytes of data takes time.

In order to aid investigators in this process, various tools have been developed for use in forensic computing. These tools help by providing automated searching for key attributes. For example, a tool could be used to discover all of the image files on a computer or be used to search for evidence of encryption software or for files whose extension does not match the file header. Tools increase the efficiency of a search and decrease the time needed to complete a full forensic analysis of a computer. It will always be more efficient to search the contents of a device using a tool than for a human to look at it file by file simply because computers are capable of processing information much faster than humans.

Various methods are implemented in tools for searching. Some of these methods are more efficient or quicker than others. The questions of how to search mass amounts of data is an ongoing one, with continuing research. It is proposed here that using ontologies as filtering mechanisms may prove effective.

In this paper, we discuss the effectiveness of using ontologies for searching within a forensic computing context. An ontology is a specification about an aspect of the world. The effectiveness of ontologies for searching the World Wide Web continues to be researched, as does the effectiveness of ontologies in knowledge management but no previous work into the use of ontologies in forensic computing was discovered.

## 2. EXAMINING ELECTRONIC EVIDENCE

The process of "acquiring, preserving, retrieving, and presenting data that has been processed electronically and stored on computer media" (Noblett et al., 2000) is known as computer forensics. Here we also use the term electronic evidence which has been in long term use by Law Enforcement and other agencies and which is here seen to subsume, and include, terms such as data forensics, digital forensics, forensic computing and network forensics.

In Australia, both Law Enforcement and academia rely on the definition of Forensic Computing first proposed by McKemmish (1999) who stated that forensic computing is the "process of identifying, preserving, analysing and presenting digital evidence in a manner that is legally acceptable". This new science of computer forensics evolved to follow the basic methodologies of other forensic sciences. Rules of evidence vary from country to country and even from state to state but certain aspects remain the same, however. McKemmish (1999) defines the four key elements of forensic computing as identification, preservation, analysis and presentation of electronic evidence.

Analysis is perhaps the most important stage of a computer forensic investigation. It involves extracting electronic evidence and processing and interpreting the data for presentation in court.

### 2.1 Sources of Electronic Evidence

Digital evidence can be gathered from any data storing electronic device. People most associate computer forensics with PCs and laptop computers, but it also applies to mobile phones and PDAs, as well as other electronic storage devices. Hard drives are the most common type of large storage media in a PC (or laptop computer) that are examined. Storage media such as floppy discs, CDs and DVDs are also examined for digital evidence. As such, the formats in which digital evidence can be found are many and varied.

The size of storage media has also increased dramatically in recent years, with hard drives offering more than 60 GB of storage and DVDs approximately 4.5 GB. This presents further challenges to computing forensics due to the sheer amount of data that must be sifted through in order to find the relevant evidence. A single CD is capable of storing hundreds or even thousands of files. Viewing and analysing each and every one of such files would be time consuming and inefficient, as many of the files would not be related to what was being investigated.

In an unpublished keynote at the IFIP 11.9 Working Group Annual Meeting in Orlando, Florida in February 2006, Professor Gene Spafford of Purdue University noted that in 2000, the entire world's collection of electronic data was approximately 200TB but that in 2006 this much data can be stored on a roomful of powerful computers. Hence, one of the greatest challenges for Law Enforcement is the timely and targeted discovery and acquisition of an exponentially-increasing quantity of electronic evidence which may be hidden within the vast quantities of electronic data stored within an organisation under investigation.

## 3. FILTERING

In order to gather the relevant electronic evidence in a reasonable period of time, some kind of filtering is needed. Searching through many thousands of MBs of data takes time and this time is just not available for an investigation. As Noblett et al. (2000) state, time limitations and judicial constraints may not permit the examiner to search every file. Anderson (2001) concurs with this, adding that the "US court system expects legal discovery to be conducted in a 'reasonable' period of time" and that judges and lawyers do not generally understand the amount of data that must be sorted through in order to find the relevant evidentiary data. As such investigators need to focus on the relevant data (the data which is of interest in the investigation as opposed to that data which is simply extraneous and unnecessary). Filtering allows a search to be focused and more relevant. This does require that the examiner be aware of "specific knowledge of investigative details" (Noblett et al., 2000), that they know what type of data they are looking for, be it e-mails or graphical files.

Tools are currently available to perform various types of forensic searching. They focus on attributes such as files types and encryption. In this way the search space can be reduced. For example, rather than examining every file on a storage medium only those with relevant file types can be looked at. Looking merely for file extensions for determining its type is insufficient, however. "Any user or application can change a file's name and extension at any time" (McDaniel & Heydari, 2002). A further example of how filtering can be used is for the identification of encryption software and/or encrypted files. Compressed files can also be targeted.

The purpose of filtering data is to reduce the search space into something more manageable – "to partition the device into logical areas following a specific objective" (Slay & Jorgensen, 2005). This can be done not only by eliminating various types of files from the search but also by eliminating various sections of the drive from the search. For example, the Windows folder on a PC or laptop running any version of Microsoft Windows is unlikely to be used for storage by any average computer user as it contains files pertaining to the operating system itself. Certain areas can also be targeted. Hama and Pollitt (1996) maintain that "reviewing directory listings is still a valuable technique".

The average user, for example, would organise their directories and files in a manner similar to the way they would organise a physical file cabinet – orderly with accurate labels so things can be identified. A user who is not expecting a forensic investigation of their system would also be inclined to organise their files with an eye towards convenience, rather than secrecy. However, a user who is more experienced or aware of the fact that they have engaged in illegal activities or have illegal files on their computer is more likely to have attempted to conceal them. However, as Hama and Pollitt (1996) note, the majority of users do not view their electronic files as evidence against themselves. Slay & Jorgensen (2005) disagree with this, however, saying instead that "the suspect under investigation may have gone to some trouble to deliberately hide the information that could incriminate them". This more modern perspective is more likely to be correct, as users today are much more computer literate than in the past and their awareness of what is and isn't legal is higher, perhaps due to many recent highly publicised court cases.

Filters are a valuable tool in reducing search space but they can also be used for specific purposes, such as looking for a certain type of file (Slay & Jorgensen, 2005). Such filters can be incorporated in tools designed to aid investigators. A tool that could be used on a suspect's computer at the scene of an investigation could be able to identify possibly suspicious data and/or files if such filters were incorporated. Such a pre-analysis could aid investigators in deciding which devices should be high priority for a full forensic analysis at the laboratory. This is especially important as the capacity of various devices increases.

Slay & Jorgensen (2005) indicate that filters can be divided into four main categories namely Inclusion (AND/OR), Exclusion (NOT), Grouped and Isolated (COMBINATIONS). This division is based on the "standard mathematical discrete logic principles". These filters are defined by their purpose.

- An inclusion filter includes something. For example, an inclusion filter could be designed to include all jpeg format image files in its end result.

- An exclusion filter excludes something. For example, an exclusion filter could be designed to exclude all image files that are less than 100kb in size.

- A grouped filter is designed to find similar chunks of data that are located near one another; it looks at groups of similar data. For example, if a directory contains over 90% of the same file type or if two directories near one another contain the same file types.

- An isolated filter looks for data that is dissimilar to the data around it. For example, a single image file in a directory that otherwise contains only PDF files.

Slay & Jorgensen (2005) used multiple filters together through the use of intersection. Using intersection rules to join filter types together allows four different outcomes – "Accepting everything each filter returns, accepting everything except where the filters intersect, accepting only the intersection, and accepting only the results of one filter plus or minus the intersection with another filter".

Using these various filters in combination resulted in a framework that was implemented in a simple hard disk analysis tool. However the search method used was a simple linear search which we find implemented in some common commercial software, such as Encase. In this case we find that the data is being searched repetitively in the process of finding, in this implementation, files with particular extensions.

While this research was an examination of the use of filters for searching for file extensions, this implementation drew on the most common method used for **keyword** searching, or in fact for searching for any **key attributes** within the image such as file extension, file content.

This method involves identifying a single key attribute which potentially needs to be identified within a disc image, and then programming the forensic tool to search for each incidence of the key attribute within the hard disk image. However, this is very time consuming, since the search tool needs to read all the data, and then search for the key attribute.

An alternative method currently used for **keyword** searching is to index the data first and then use the index automatically to provide the locations of the keywords within the image when the software is run (this is the method implemented in a product such as Access Data's Forensic Tool Kit). This is also a very time consuming process but does ensure that all text on a given image is indexed and then all keywords may be found very quickly.

However, other more efficient and quicker means of searching need to be explored to deal with the exponential growth in electronic evidence which needs investigation. In current practice a typical Law Enforcement investigator may need to wait 15 hours for one simple keyword search of a hard disk image. This produces an almost unmanageable wait time and has enormous workload implications; other techniques need to be sought to search hard disk images.

## 4. ONTOLOGIES

Ontologies were initially a concept of philosophy (Zúñiga, 2001) that were adapted for use within the realm of information systems. Ontologies are commonly used in artificial intelligence and in knowledge management. It is proposed that ontologies may aid in filtering mass amounts of data.

The commonly cited definition of an ontology is by Gruber (1993) who says an ontology is "an explicit specification of a conceptualisation". This definition was quoted by Ding and Foo (2002) who also declared that the term ontology "refers to that shared understanding of some domain of interest". Ding and Foo (2002) go on to state that ontologies are "often conceived as a set of classes, concepts, relations, functions, axioms and instances". This arises in the work of Gruber (1993) who declares that a program's ontology can be described by defining the various objects that exist within the programs' domain. Such definitions associate the names of the various objects with text that describes what names mean in a manner understandable by humans.

Noy and McGuiness (2001) define an ontology as a "formal explicit description of concepts in a domain of discourse (classes or concepts), properties of each concept describing various features and attributes of the concept (slots, roles or properties) and the restrictions on slots (facets or slot restrictions)". They go on to say that a knowledge base is composed of an ontology and a set of individual instances of classes.

An ontology is a representation of some aspect of the world. An ontology defines all the different things that make up that aspect and how they relate to each other. Defining a domain and the objects within that domain is the core of an ontology. Classes are the key to this.

Classes are used to represent the objects in a domain. Classes are then organised into a hierarchy that defines the relationship between them. The "Is-A" relationship is the primary one in most ontologies (Lammari & Métais, 2004), allowing classes to be structured easily into hierarchies.

For example, an ontology of programming languages would state that Java is a programming language, giving rise to the following structure where the Java object is represented as a subclass of Programming Languages:

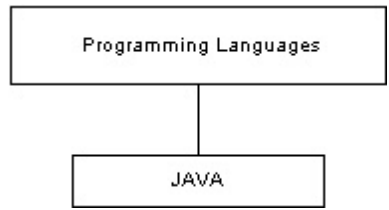

Figure 1 – Simple Programming Language Ontology

In further refinement we could add that Java is an object oriented language:

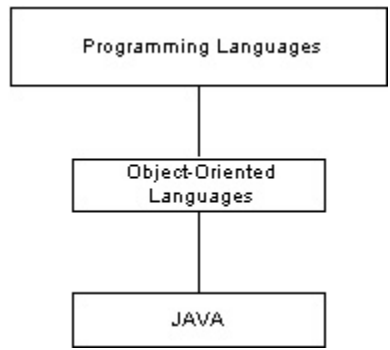

Figure 2 – Refined Simple Programming Language Ontology

C++ is also object oriented, but C is not, so:

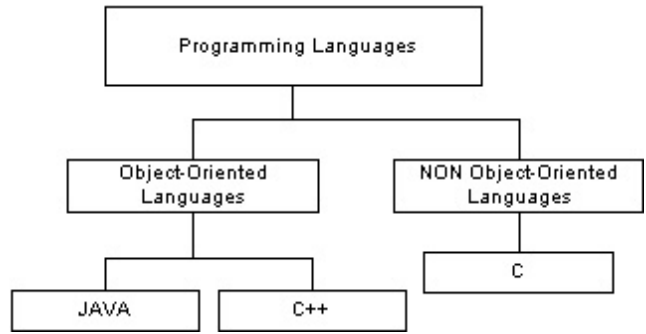

Figure 3 – Evolved Simple Programming Language Ontology

This very basic hierarchy quickly and easily shows the most basic relationship between these objects, forming a small portion of Programming Languages ontology. This ontology was drawn in the form of a hypertree, presented as a tool for visualising ontologies by Souza et al. (2003).

The hierarchy of an ontology must be accurate, as the ontology needs to accurately represent something from the world. Multiple inheritance may sometimes be necessary. Noy & McGuinness (2001) provide some basic questions for use in defining an ontology.

The domain of an ontology is to be considered first. What is the domain is the first question to be asked and the scope of the domain determined. For the domain of Programming Languages, for example, the scope might need to be narrowed from the domain of all programming languages to only those programming languages that are object oriented. Alternatively, it could be limited to only include programming languages invented after a certain date.

The second of Noy and McGuinness's (2001) questions requires that the purpose of an ontology be looked at, that creators ask what the ontology is to be used for. Asking this question aids in determining the scope of the domain. A purpose driven ontology can discard unnecessary aspects of a domain. For example, if the programming languages ontology is to be used in a course database at a university then languages not taught at the university need not be included.

The third question to be asked is actually a series of questions, asking what types of questions the ontology should be able to answer. An ontology is to be a representation of some part of the world, so it must be able to answer questions that would normally be asked in regards to the ontology's domain. For this reason, the ontology should be thoroughly questioned to ensure it's accuracy in modelling its domain. A programming language ontology could need to answer questions in regards to whether or not a language is object oriented, when a language was created, or what universities teach that language.

The users of an ontology are the focus of the fourth of Noy and McGuinness's (2001) questions. The users of an ontology also affect its design. For example, students at a university might wish an ontology of Programming Languages that the university teaches to include information regarding fees and textbooks. And ontology must also be maintained over its lifetime. The programming languages that a university teaches may change and the ontology would also need to change. The person maintaining an ontology may not describe it in exactly the same words or concepts as the users so some sort of terminology mapping may be necessary.

Noy and McGuiness (2001) emphasise that developing an ontology is by necessity an iterative process. They also declare that modelling of a domain can be done in many ways and that there is "no one correct way to model a domain". They further recommend that the concepts within an ontology be "As close to objects (physical or logical) and relationships in your domain". It is likely that when an ontology's domain is described the nouns in sentences will be able to be immediately mapped as objects and the verbs as relationships.

Defining the hierarchy of an ontology can be done in several different ways. Noy and McGuinness (2001) present three different high-level approaches whereas Lammari and Métais (2004) present algorithms for use in creating an ontology.

Noy and McGuinness's (2001) first approach is a "top-down development process". This involves defining the most abstract concepts to be included in the ontology and then becoming more specific. For example, starting with the concept of Programming Languages and then specifying subclasses such as Java and C++.

The second of Noy and McGuinness's (2001) approaches is a "bottom-up development process" where the subclasses/objects are defined first and are then grouped together into the higher level classes. For example, starting with the concepts of Java and C++ and then grouping them together as Programming Languages.

The third approach of Noy and McGuinness (2001) is a "combination development process" where the top-down and bottom-up approaches are used together. For example, starting with the high-level concept of Programming Languages and then using a top-down approach to add the concepts of Java and C++ to the ontology. Java and C++ can then be grouped together as object oriented programming languages in a bottom-up approach. The two approaches can then continue to be used in combination until a complete ontology has been developed.

Noy and McGuinness's (2001) approach is high-level and allows a creator to both create and alter an ontology in a very abstract manner. Lammari and Métais (2004) present a more specific, systematic method for creating an ontology, providing algorithms which can aid a creator in the creation of a complete and accurate ontology.

Lammari and Métais (2004) base their algorithms on "existence constraints". Various existence constraints are used to describe the relationships and set of keywords are used to specify the grouping that will form the hierarchy of the ontology. This process is much stricter than the conceptual one of Noy and McGuinness (2001).

Ontologies are "an important emerging discipline" (Ding and Foo, 2002). Research in the use of ontologies is ongoing as their popularity increases. A majority of this research is in the area of organising and searching the World Wide Web, as discussed by Mahalingam and Huhns (1997) and also by Stojanovic (2005), and in the area of extracting data from databases, as referred to by Biskup and Embly (2003) and also by Gruenwald, McNutt and Mercier (2003). The use of ontologies in the arena of forensic computing is new but it is potentially powerful, given the success of ontologies in other arenas.

## 5. METHODOLOGY

In order to research the effectiveness of using an ontology for filtering data in a forensic computing context a sample ontology has been developed and evaluated. The sample ontology is relatively small, designed for a specific purpose. The sample ontology is, however, sufficient to evaluate the effectiveness of ontologies in forensic data searching.

The sample ontology's purpose is to identify possibly suspicious data on a computer. This ontology is to be implemented in a pre-analysis triage tool being developed for the South Australian Police. The purpose of this tool is for a detective to be able to use at the scene for a five minute scan of a suspect's computer in order to look for possibly suspicious data. The tool will enable investigators to decide whether or not a full forensic analysis of a device is a high-priority.

The term possibly suspicious data is used in this paper to describe data that may indicate wrongdoing on the part of a suspect. A tool can not make the determination of whether or not that data it discovers are indeed suspicious, hence the use of the term possibly suspicious. The data discovered by a tool must then be examined by an investigator to determine whether or not it is in fact suspicious. For example, a tool may find a large amount of image files but only an investigator can determine if the images are child pornography or merely photos of family members or landscapes taken with a digital camera. While a tool can locate possibly suspicious data only an investigator can interpret it.

The particular tool in which an ontology was developed for the purpose of the research presented in this document is a pre-analysis tool. It is to be used at the scene for triage before a full forensic analysis is undertaken in a laboratory. Its purpose is to allow investigators to determine if a device should be placed at high priority for a full forensic investigation and analysis. With many people owning multiple high capacity devices this may be critical as investigations must be completed within a specific timeframe. Full analysis of all devices may not be possible due to time constraints; hence the need for pre-analysis tools to aid in determining the priority of a forensic analysis.

### 5.1 ZSAT Tool

The tool within which an ontology was developed in order to evaluate the effectiveness of using ontologies in computer forensic searching is Version 2 of the Zero Skills Analysis Tool (ZSAT). The ZSAT is being developed by UniSA for SAPOL. Its purpose is to allow a detective at the scene to scan a computer owned or used by a suspect for possibly suspicious data. The ZSAT's primary requirements are that it not alter any data on a suspect's computer, ease of use, and also that it either complete or halt at 5 minutes. This means that the search must be both efficient and effective. It is proposed that the use of an ontology to filter data will achieve this. To test this, an ontology was implemented as the filtering mechanism for the ZSAT.

ZSAT is developed for use on computers running Microsoft Windows XP (Home or Pro). The ZSAT is being implemented as a stand alone program that can be inserted into a computer as a boot disc. In this way the computers OS will never boot, preventing it from changing files. The ZSAT itself will not alter any files, thus complying with the rules of evidence.

The ZSAT has a graphical user interface (GUI, designed for ease of use, allowing any police officer to execute the search, rather than requiring a computer forensics expert be present. It would be ineffective for computer forensics experts to be called out to every scene that included an electronic device. As such, a default search is available. This can be altered to target different file types by more experienced users who are aware of the context of the crime being investigated. A music and piracy investigation would wish to look first for music and video files for example, while an investigation into child pornography would be primarily interested in image and video files. Targeting specific types of files increases the likelihood of success of a search. This is particularly important in a pre-analysis tool such at the ZSAT as it will be used to determine the priority of full investigations of electronic devices.

### 5.2 Purpose of the Ontology

An ontology was implemented as the filtering mechanism within the ZSAT. This ontology is purpose-driven, designed specifically for this pre-analysis tool. As such its purpose is to filter the data on a suspect's computer in order to locate possibly suspicious data. This must be done as quickly and efficiently as possible because of the limited five minute time in which the ZSAT's search must operate.

It is proposed that an ontology will filter data more efficiently than traditional filtering methods. One easy filtering mechanism is to simply compare every file against the desired types, as depicted in figure 4. This linear searching is both time consuming and inefficient.

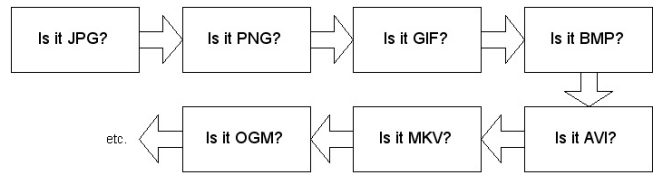

Figure 4 – Linear Searching

Such a search is targeted in that it searches for specific types of file but its effectiveness is limited. Its limitations become even more obvious when one begins to want to not only find files of specific types but also to decide which of them may be suspicious and which a probably innocent. This paper attempts to answer the question of whether or not an ontology will offer a more efficient filtering mechanism. This will be done firstly by the implementation of an ontology in a small pre-analysis tool and, in further work, by comparing the performance of this ontological program with a non-ontological implementation of the same program.

### 5.3 Design of the Ontology

An ontology defines the various objects within a program's domain. It can be said that an ontology defines the world as a program sees it. In this case, the ontology is to act as the filtering mechanism for mass amounts of data, so its domain will encompass and define the file types being targeted as possibly suspicious. The domain is at the same time narrowed to only include the desired files types, ignoring all others.

In this instance the ZSAT is required to look for image and video files. Music files are also an option in executing the ZSAT and as such are included in the hypertrees used to represent this ontology although by default they are disabled during the search. Music files have been disabled in the default search of the ZSAT because while piracy of music will be of concern to police in the future at this point in time it is beyond the resources of the department.

This ontology was designed primarily using the top-down development process conceptualised by Noy and McGuiness (2001) described above. This process involved an iterative process of determining and defining the objects within the domain and representing them as classes. It began with the top level concept of required file types, the file formats the ZSAT is required to find and assess as potentially suspicious. This initial stage is shown in Figure 5.

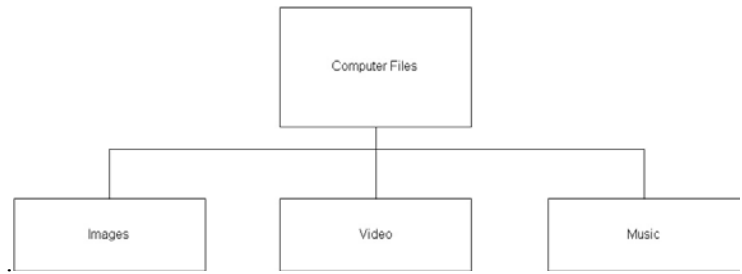

Figure 5 – Ontology, Evolution 01

This top level concept then became more specific, the domain limited to only include Image, Video, and Music files. Specific formats of these generalised file types were then specified, defining the world as the ontology sees it.

Each file type concept in the ontology is a class, with a hierarchy of "is-a" relationships that is easily understood. Figure 6 shows the first evolution of an ontology for use within the ZSAT. It is drawn in the form of a hypertree, presented as a tool for visualising ontologies by Souza, Santos, and Evangelista (2003). The hypertree representation makes it easy to see the relationships between objects.

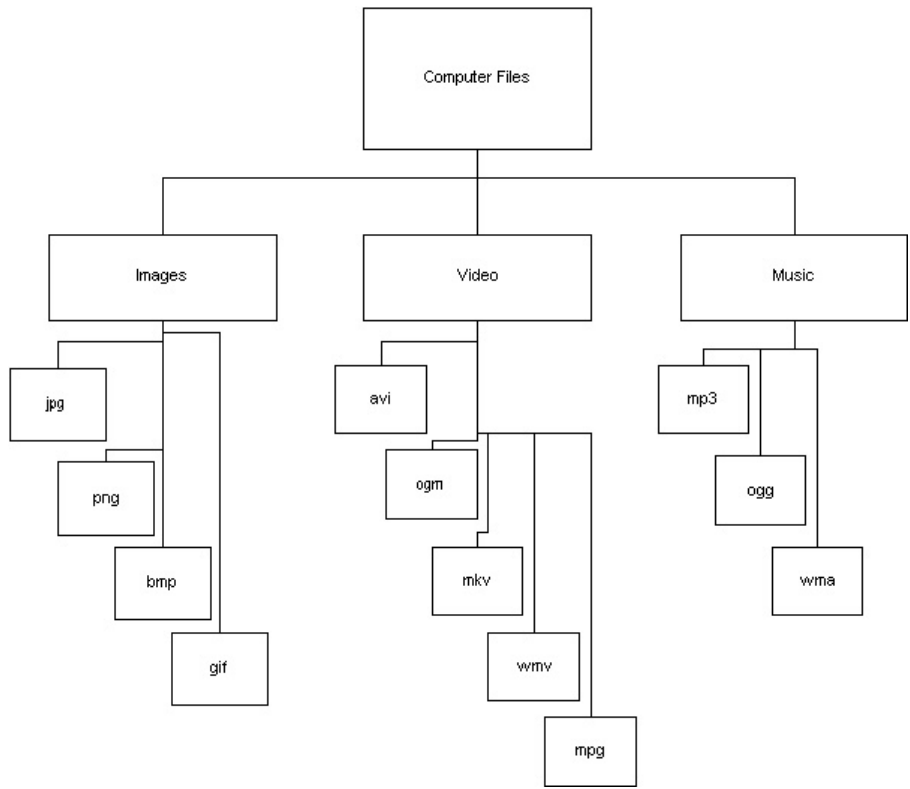

Figure 6 – Ontology, Evolution 02

It should be remembered that the users and purpose of an ontology affect its design. This was certainly true in the creation of this ontology for the ZSAT. Simplicity was paramount whilst still remaining effective and efficient. As advised by Noy & McGuiness (2001) the concepts that make up the various classes in this ontology were modelled on the objects required within the domain the ontology encompasses

This initial ontology easily narrowed the domain of the ontology from that of all computer files to only those the ZSAT is required to search for – Images, Video and Music files. Specific formats of these file types were then defined in the ontology, those that were considered to be more likely to contain suspicious data. The ontology is unaware that other formats of these file types even exist; its domain is narrowed to only include what is required. File formats were chosen based on those most commonly used, particularly on the Internet, both now and in previous years. The native formats of items such as digital cameras were also considered in choosing these file formats.

Each sub-domain of file types within the ontology of all required file types has its own file formats which are required to be examined as being possibly suspicious. These file formats make up sub-domains of their own, each format with unique characteristics.

The ontology shown in Figure 6 shows the file types that make up the ZSAT's ontology. A further iteration was then required in order to determine what made each type of file suspicious. The aspects of an image file which make it suspicious are different to those which make a video file suspicious, for example. The file size which makes an image may not correspond to a suspicious video file size. For example, an image file that is 200kb may be considered possibly suspicious, but a 200kb video file is not. The ontology shown in Figure 8 was finally decided upon, but not before another option was considered.

This alternative possible structure for the ontology was the placing Suspicious as a subclass of each of the file types being searched for (i.e. – as a subclass of Images, for example). This makes it a sibling class of each of the specific types of files being searched for (such as jpg, png, etc.). This is shown in Figure 7. This was initially considered as it was originally assumed that the factors that make one specific type of file suspicious would be the same for the remaining sibling classes of that file type. It could have been declared as a property of each file type in this design rather than as a subclass, but this was considered to be considerably less efficient if the property was the same for each of these sibling classes.

The realisation that the factors that make one specific format of a file type suspicious were not the same as those of its sibling classes caused a further refinement of the ontology. The difference in what makes specific formats of file suspicious can be demonstrated within the Images domain. The factors that make a bitmap suspicious are different to those that make a jpeg suspicious due to the differing compression used in each format. A bitmap that is possibly suspicious will be much larger than a possibly suspicious jpeg. Due to these realised differences the ontology evolved into the one seen in Figure 8.

The ontology shown in Figure 8 completely defines the required domain of possibly suspicious files that the ZSAT program is required to search for and filter. The suspicious subclass of each image type then contains the various properties that contribute to a file being considered possibly suspicious by the ZSAT. These properties include file size, keywords in the file name, and file location.

The hierarchical structure of the designed ontology is easily translated into classes, giving an immediate hierarchical structure to the code as well. This hierarchical ontology can also be shown in a hierarchy, as shown in Figure 9.

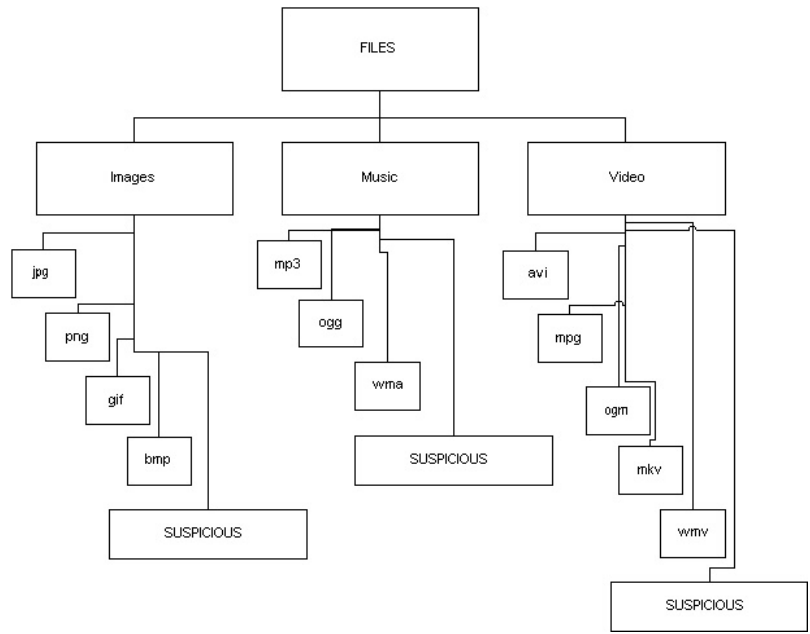

Figure 7 – Considered Alternative Ontology

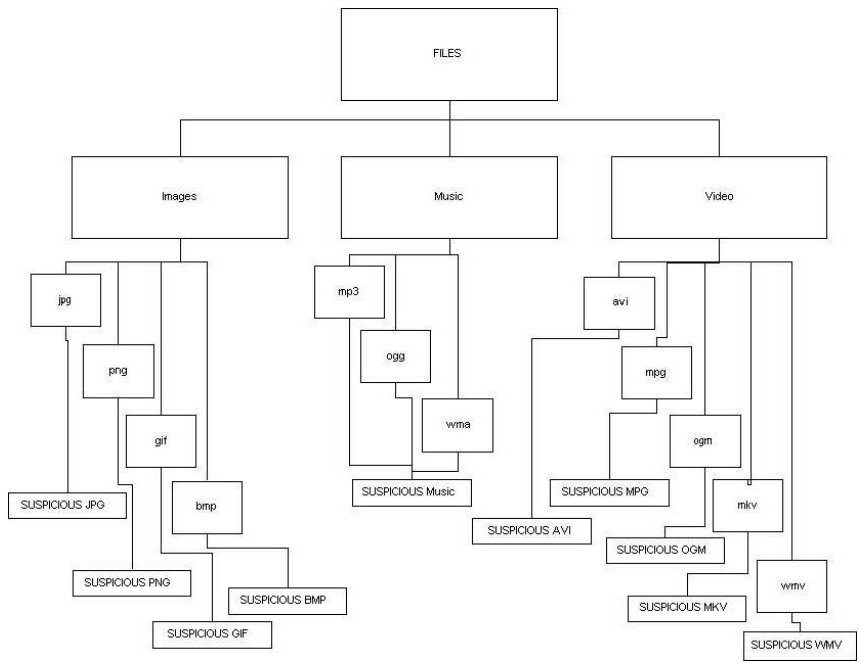

Figure 8 – Final Ontology

```
|-Files
|- -Images
|- - -jpg
|- - -png
|- - -gif
|- - -bmp
|- -Video
|- - -avi
|- - -wmv
|- - -mkv
|- - -mpeg
|- - -ogm
|--Music
|- - -mp3
|- - -wma
|- - - ogg
```

Figure 9 – Hierarchy View of Ontology

## 6. RESULTS

The program was written using Java. This created various problems in itself as it required the creation of a file object for each file being examined by the program. This was due to various implementations in the standard Java libraries. It is believed that implementing a hierarchical ontology for file searching and filtering in another programming language such a C or C++ would prove to be more effective.

The program was tested on an 80GB hard drive containing a total of 173,728 files in 10,808 folders. These files take up 50.1GB of the drive, leaving 24.3 GB free. This drive contains at least 500 images in various locations and 10 video files that should be considered suspicious by the program due to file size alone.

On the most basic level of search, simply finding the required file formats with no filtering for what is considered to be possibly suspicious, took 126531 milliseconds to complete. Upon implementing filtering within the program, the ZSAT search completed in 195621 milliseconds. Filtering was implemented based on file location, proximity to other files of the same type, file size, and keywords in the file name. These results clearly show the potential of ontologies to quickly select priority targets for forensic analysis.

## 7. CONCLUSION AND FURTHER WORK

It is the purpose of this paper to evaluate the effectiveness of using ontologies for searching within a forensic computing context. An ontology is a specification about an aspect of the world. A small sample hierarchical ontology was created specifically to search for images and video files and was then compared to a program which performed a linear search. The ontology proved successful in finding files and in filtering the results within a suitable timeframe.

The small sample ontology tested for the purposes of this document demonstrates the power of ontologies for use in searching and filtering mass amounts of data in a forensic computing context. It can be concluded that the use of ontologies will indeed aid investigators in filtering mass amounts of data and thus also in completing investigations within the required timeframe.

The growth in research into the semantic web and the success of ontologies within the domain of searching the World Wide Web suggest that ontologies would also be successful in searching data within a forensic computing context. The research contained within this document shows this to be true; ontologies can be used in forensic computing searching tools when data filtering is required. Further work is required to discover their true potential, however.

Further evaluation of ontologies for use in filtering large amounts of data within a forensic computing context is required before any real conclusion concerning their effectiveness can be reached. Larger ontologies involving larger domains and more sub domains would need to be tested before the true effectiveness of ontologies can be seen.

Other questions that were raised during this research include the need for the creation of a standard method or procedure for the creation and implementation of ontologies within a forensic computing, and even in a wider context, as there does not currently appear to be one. This could lead to problems in the future if ontologies are used in the creation of forensic computing tools. Indeed the lack of global and national standard procedures and guiding principles is one of the challenges that face the entire discipline of forensic computing in regards to producing admissible and credible evidence.

## 9. AUTHOR BIOGRAPHIES

**Dr. Jill Slay** holds degrees in Mechanical Engineering, applied computing and further education and a PhD from Curtin University of Technology. She is a Fellow of the Australian Computer Society, MIEEE and is a Certified Information Systems Security Professional. She leads the Enterprise Security Management Laboratory in the Advanced Computing Research Centre at the University of South Australia Jill has published widely in IA and forensic computing. Currently, she carries out collaborative research in Forensic Computing and IT Security with industry and government partners in Australia and the USA.

**Fiona Schulz** is a recent graduate in Software Engineering from the University of South Australia and now works for the Australian Department of Defence.